\documentclass[10pt,conference]{IEEEtran}
\usepackage{amstext,url,boldmath,amsmath}

\def\Magma{{\textsf{Magma}}}
\def\QECC(#1,#2,#3,#4){[\![#1,#2,#3]\!]_{#4}}
\def\tr{\mathop{\rm tr}\nolimits}

\newtheorem{theorem}{Theorem}
\newtheorem{corollary}[theorem]{Corollary}

\newtheorem{lemma}[theorem]{Lemma}

\begin{document}

\title{Quantum Block and Convolutional Codes from Self-orthogonal
  Product Codes}

\author{\authorblockN{Markus Grassl}
\authorblockA{
Institut f\"ur Algorithmen und Kognitive Systeme\\
Arbeitsgruppe Quantum Computing\\
Fakult\"at f\"ur Informatik, Universit\"at Karlsruhe (TH)\\
Am Fasanengarten 5, 76\,128 Karlsruhe, Germany\\
Email: grassl@ira.uka.de}
\and
\authorblockN{Martin R\"otteler}
\authorblockA{
NEC Labs America, Inc.\\
4 Independence Way\\
Princeton, NJ 08540, USA\\
Email: mroetteler@nec-labs.com
}
}

% avoiding spaces at the end of the author lines is not a problem with
% conference papers because we don't use \thanks or \IEEEmembership
%Homer Simpson\authorrefmark{2},
%Montgomery Scott\authorrefmark{3} and
%Eldon Tyrell\authorrefmark{4}}
%Telephone: (800) 555--1212, Fax: (888) 555--1212}
%\authorblockA{\authorrefmark{4}Tyrell Inc., 123 Replicant Street, Los Angeles, California 90210--4321}}

\maketitle
\pagestyle{plain}

\begin{abstract}
We present a construction of self-orthogonal codes using product
codes.  From the resulting codes, one can construct both block quantum
error-correcting codes and quantum convolutional codes.  
We show that from the examples of convolutional codes found, we can
derive ordinary quantum error-correcting codes using tail-biting with
parameters $\QECC(42N,24N,3,2)$. While it is known that the product
construction cannot improve the rate in the classical case, we show
that this can happen for quantum codes: we show that a code
$\QECC(15,7,3,2)$ is obtained by the product of a code
$\QECC(5,1,3,2)$ with a suitable code.
\end{abstract}

\section{Introduction}
Quantum convolutional codes are motivated by their classical
counterparts \cite{Bla03}. As in the classical case the idea is to
allow for the protection of arbitrary long streams of information in
such a way that as many errors as possible can be corrected.  To
achieve this the information is ``smeared out'' to the output stream
by adding a certain amount of redundancy, but at the same time meeting
the requirement to be local, {\em i.\,e.}, encoding/decoding can be done by
a processes which needs only a constant amount of memory. In
\cite{OlTi04} the basic theory of quantum convolutional codes has been
developed. There it has been shown that, similar to the classical
codes, quantum convolutional codes can be decoded by a maximum
likelihood error estimation algorithm which has linear
complexity. However, the authors only gave an example of one (rate
$1/5$) quantum convolutional code.  This research was motivated by the
question to find new examples of quantum convolutional codes. The
construction presented in this paper resorts on the idea of product
codes. An extra requirement imposed by the applicability to quantum
codes is that the dual distance has to be high. The main source of the
examples presented at the end of the paper are two-dimensional cyclic
codes (sometimes also called ``bicyclic codes''). We apply this to the
situation where the code is a product code of two Reed-Solomon codes.

\section{Self-orthogonal Product Codes}\label{sec:inner_products}
\subsection{Quantum error-correcting codes from classical codes}
Most of the constructions for quantum error-correcting codes (QECCs)
for a quantum system of dimension $q$ (\emph{qudits}), where
$q=p^\ell$ is a prime power, are based on classical error-correcting
codes over $GF(q)$ or $GF(q^2)$.  The so-called CSS codes (see
\cite{CaSh96,Ste96:error}) are based on linear codes $C_1$ and $C_2$
over $GF(q)$ with $C_2^\bot\subseteq C_1$.  Here $C_2^\bot$ is the
dual code of $C_2$ with respect to the Euclidean inner product.  In
particular, if $C=C_1=C_2$ this implies that $C^\bot$ is a weakly
self-dual code.  The construction can be summarized as follows:
\begin{lemma}\label{lemma:CSS}
Let $C=[n,k,d]_q$ be a weakly self-dual linear code, {\em i.\,e.},
$C\subseteq C^\bot=[n,n-k,d^\bot]_q$.  Then a quantum error-correcting
code encoding $n-2k$ qudits using $n$ qudits, denoted by
${\cal C}=\QECC(n,n-2k,d_q\ge d^\bot,q)$ exists.
\end{lemma}
Another class of quantum codes can be obtained from codes over
$GF(q^2)$ which are self-orthogonal with respect to the Hermitian
inner product, denoted by $C\subseteq C^*$.  Both cases can be
generalized to a construction of QECCs based on additive codes over
$GF(q^2)$ which are self-orthogonal with respect to the symplectic
(trace) inner product, {\em i.\,e.} $C\subseteq C^\star$ \cite{AsKn01}.

\subsection{Inner products on vector spaces over $GF(q)$ and $GF(q^2)$}
In this paper, we will use three different inner products on vector
spaces over $GF(q)$ and $GF(q^2)$ which are defined as follows:
\begin{alignat}{2}
\noalign{Euclidean:}
\bm{v}\cdot\bm{w}&:=\sum_{i=1}^n v_i w_i\quad\text{for
  $\bm{v},\bm{w}\in GF(q)^n$}\label{eq:Euclidean}\\
\noalign{Hermitian:}
\bm{v}*\bm{w}&:=\sum_{i=1}^n v_i w_i^q\quad\text{for
  $\bm{v},\bm{w}\in GF(q^2)^n$}\label{eq:Hermitian}\\
\noalign{symplectic:}
\bm{v}\star\bm{w}&:=\sum_{i=1}^n \tr(v_i w_i^q)\quad\text{for
  $\bm{v},\bm{w}\in GF(q^2)^n$}\label{eq:symplectic},
\end{alignat}
where $\tr(x)$ denotes the trace of $GF(q^2)$ over its prime field
$GF(p)$.  Both the Euclidean and the Hermitian inner product are
bilinear over $GF(q)$ respectively $GF(q^2)$, but the symplectic inner
product is only $GF(p)$-bilinear because of the trace map.  For codes
which are linear over $GF(q)$, linear over $GF(p^2)$, or additive
({\em i.\,e.} $GF(p)$-linear), one can define a dual code with respect to
the inner products (\ref{eq:Euclidean}), (\ref{eq:Hermitian}), or
(\ref{eq:symplectic}), respectively.  The three cases are summarized
in Table~\ref{table:products}.

\begin{table}
\caption{Notation used for the three different inner products and the
  corresponding dual codes.\label{table:products}}
\centerline{\normalsize
\begin{tabular}{l|c|c|c}
& dual code &inner product & linear over\\
\hline
Euclidean  & $C^\bot$  & $\bm{v}\cdot\bm{w}$ & $GF(q)$  \\
Hermitian  & $C^*$     & $\bm{v}*\bm{w}$     & $GF(q^2)$\\
symplectic & $C^\star$ & $\bm{v}\star\bm{w}$ & $GF(p)$  
\end{tabular}
}
\end{table}

Next, we consider inner products on tensor products of vector spaces.
\begin{lemma}\label{lemma:linear_tensor}
For all $\bm{v},\bm{v}'\in GF(q)^n$ and $\bm{w},\bm{w}'\in GF(q)^m$,
we have
\begin{equation}
(\bm{v}\otimes\bm{w})\cdot(\bm{v}'\otimes\bm{w}')=
(\bm{v}\cdot\bm{v}')(\bm{w}\cdot\bm{w}'),
\end{equation}
{\em i.\,e.}, the Euclidean inner product is compatible with the tensor
product of vector spaces over $GF(q)$.  Furthermore, for all
$\bm{v},\bm{v}'\in GF(q^2)^n$ and $\bm{w},\bm{w}'\in GF(q^2)^m$, we
have
\begin{equation}
(\bm{v}\otimes\bm{w})*(\bm{v}'\otimes\bm{w}')=
(\bm{v}*\bm{v}')(\bm{w}*\bm{w}'),
\end{equation}
{\em i.\,e.}, the Hermitian inner product is compatible with the tensor
product of vector spaces over $GF(q^2)$.
\end{lemma}
\begin{proof}
The tensor product of two vectors is given by
$(\bm{v}\otimes\bm{w})=(v_iw_j)_{i,j}$. Then for the Euclidean inner
product we get
\begin{alignat*}{2}
(\bm{v}\otimes\bm{w})\cdot(\bm{v}'\otimes\bm{w}')\kern-25mm\\
&=\sum_{i=1}^n\sum_{j=1}^m v_iw_j v'_i w'_j
=\Bigl(\sum_{i=1}^n v_i v'_i\Bigr)\Bigl(\sum_{j=1}^m w_j w'_j\Bigr)\\
&=(\bm{v}\cdot\bm{v}')(\bm{w}\cdot\bm{w}').
\end{alignat*}
Similarly, for the Hermitian inner product we get
\begin{alignat*}{2}
(\bm{v}\otimes\bm{w})*(\bm{v}'\otimes\bm{w}')\kern-25mm\\
&=\sum_{i=1}^n\sum_{j=1}^m v_iw_j (v'_i w'_j)^q
=\Bigl(\sum_{i=1}^n v_i {v'_i}^q\Bigr)\Bigl(\sum_{j=1}^m w_j {w'_j}^q\Bigr)\\
&=(\bm{v}*\bm{v}')(\bm{w}*\bm{w}').
\end{alignat*}
\end{proof}
For the symplectic inner product, the situation is a bit more
complicated as it is only $GF(p)$-linear.  Considering $GF(q)^m$ only
as vector space over $GF(p)$, we may define the $GF(p)$ tensor product
of $V_1=GF(p)^n$ and $V_2=GF(q)^m$, denoted by $V_1\otimes_p V_2$.

\begin{lemma}\label{lemma:additive_tensor}
For all $\bm{v},\bm{v}'\in GF(p)^n$ and $\bm{w},\bm{w}'\in
GF(q)^m$, we have
$
(\bm{v}\otimes_p\bm{w})\star(\bm{v}'\otimes_p\bm{w}')=
(\bm{v}\cdot\bm{v}')(\bm{w}\star\bm{w}'),
$
{\em i.\,e.}, the symplectic inner product on the $GF(p)$ tensor product
space is the product of the Euclidean inner product on the first space
and the symplectic inner product on the second.
\end{lemma}
\begin{proof}
Similar to the proof of Lemma~\ref{lemma:linear_tensor}, we compute
\begin{alignat*}{2}
(\bm{v}\otimes\bm{w})\star(\bm{v}'\otimes\bm{w}')
&=\sum_{i=1}^n\sum_{j=1}^m \tr\left(v_iw_j (v'_i w'_j)^q\right)\\
&=\tr\left(\Bigl(\sum_{i=1}^n v_i {v'_i}^q\Bigr)\Bigl(\sum_{j=1}^m w_j {w'_j}^q\Bigr)\right).
\end{alignat*}
As $\bm{v}$ and $\bm{v}'$ are vectors over the prime field, the left
factor equals their Euclidean inner product $\bm{v}\cdot\bm{v}'$ which
takes values in $GF(p)$ only.  Using the $GF(p)$-linearity of the
trace map, the proof is completed.
\end{proof}

\subsection{Product codes}
Next we present the fundamental properties of the product of two codes
which combines two codes (see {\em e.\,g.} \cite{Bla83,MS77}).
\begin{lemma}
Let $C_1=[n_1,k_1,d_1]_q$ and $C_2=[n_2,k_2,d_2]_q$ be linear codes
over $GF(q)$ with generator matrices $G^{(1)}$ and $G^{(2)}$, respectively.
Then the product code $C_\pi:=C_1\otimes C_2$ is a linear code
$C_\pi:=[n_1 n_2,k_1 k_2, d_1 d_2]_q$  generated by the matrix
$G:=G^{(1)}\otimes G^{(2)}$, where $\otimes$ denotes 
the Kronecker product, {\em i.\,e.}
\begin{equation}\label{eq:block_structure}
G:=\left(
\begin{array}{cccc}
g_{11}^{(1)}G^{(2)} & g_{12}^{(1)}G^{(2)} & \ldots &g_{1,n_1}^{(1)} G^{(2)}\\
g_{21}^{(1)}G^{(2)} & g_{22}^{(1)}G^{(2)} & \ldots &g_{2,n_1}^{(1)} G^{(2)}\\
\vdots &\vdots&\ddots&\vdots\\
g_{k_1,1}^{(1)}G^{(2)} & g_{k_1,2}^{(1)}G^{(2)} & \ldots & g_{k_1,n_1}^{(1)} G^{(2)}\\
\end{array}
\right).
\end{equation}
If $C_1=[n_1,k_1,d_1]_p$ is a linear code over the prime field $GF(p)$
and $C_2=(n_2, p^{k_2}, d_2)_q$ is an additive code over $GF(q)$, then
$C_{\pi,p}:=C_1\otimes_p C_2$ is an additive code with parameters
$C_{\pi,p}=(n_1 n_2,p^{k_1 k_2},d_1 d_2)_q$.
\end{lemma}
The following theorem is valid for all compatible choices of inner
products on the component spaces of a tensor product space and the
tensor product space itself.
\begin{theorem}\label{theorem:product_dual}
Let $C_\pi=C_1\otimes C_2$ be the product code of the codes
$C_1=[n_1,k_1,d_1]$ and $C_2=[n_2,k_2,d_2]$.  By $H_1$ and $H_2$ we
denote generator matrices of the corresponding dual codes.
Furthermore, let $A_1$ and $A_2$ be matrices of size $k_1\times n_1$
and $k_2\times n_2$, respectively, such that the row span of the
matrices $H_1$ and $A_1$ is the full vector space and similar for
$H_2$ and $A_2$.  Then a generator matrix $H$ of the dual code of
$C_\pi$ is given by
\begin{equation}\label{eq:product_parity}
H:=\left(
\begin{array}{c}
H_1\otimes H_2\\
A_1\otimes H_2\\
H_1\otimes A_2
\end{array}
\right).
\end{equation}
\smallskip
\end{theorem}
\begin{proof}
Let $V_1$ and $V_2$ be the full vector spaces containing the codes
$C_1$ and $C_2$.  Furthermore, by $D_1$ and $D_2$ we denote the dual
code of $C_1$ and $C_2$ with respect to the inner product on $V_1$ and
$V_2$, respectively.  Using the properties of the inner products on
tensor product spaces (see Lemma~\ref{lemma:linear_tensor} and
Lemma~\ref{lemma:additive_tensor}), it is obvious that the dual code
$D_\pi$ of $C_\pi$ contains both $V_1\otimes D_2$ and $D_1\otimes
V_2$.  The intersection of these spaces is $D_0:=D_1\otimes D_2$,
spanned by $H_1\otimes H_2$. The complement of $D_0$ in $V_1\otimes
D_2$ is spanned by $A_1\otimes H_2$, and analogously for the
complement of $D_0$ in $D_1\otimes V_2$.  Hence $D_\pi$ can be
decomposed as
$$
D_\pi=\Bigl(D_1\otimes D_2 \Bigr)
   \oplus \Bigl(\langle A_1 \rangle \otimes D_2 \Bigr)
   \oplus \Bigl(D_1 \otimes \langle A_2 \rangle \Bigr).
$$
Here $\langle A \rangle$ denotes the row span of the matrix
$A$. Considering the dimension of the spaces, the result follows.
\end{proof}
\begin{corollary}
The minimum distance of the dual of the product code $C_\pi=C_1\otimes C_2$
cannot exceed the minimum of the dual distance of $C_1$ and the dual
distance of $C_2$.
\end{corollary}
\begin{proof}
The dual code $D_\pi$ of $C_\pi$ contains $V_1\otimes D_2$, {\em i.\,e.},
the product of the trivial code $[n_1,n_1,1]$ and $D_2$.  Hence the
minimum distance of $D_\pi$ cannot be larger than that of $D_2$.  The
result follows by interchanging the role of $C_1$ and $C_2$.
\end{proof}
Note that despite their poor behavior in terms of minimum distance,
the dual of product codes can be used for burst error correction (see
\cite{ChNg73,Wol65}).  For the construction of QECCs, we will make use
of the following property.
\begin{theorem}\label{theorem:tensor_selfdual}
Let $C_E\subseteq C_E^\bot$, $C_H\subseteq C_H^*$, and $C_s\subseteq
C_s^\star$ denote codes which are self-orthogonal with respect to the
inner products (\ref{eq:Euclidean}), (\ref{eq:Hermitian}), or
(\ref{eq:symplectic}), respectively.  Furthermore, let $C$ denote an
arbitrary linear code over $GF(q)$, respectively $GF(q^2)$, and let
$C_p$ be a linear code over $GF(p)$.  Then
\def\theenumi{(\roman{enumi}}
\begin{enumerate}
\item $C\otimes C_E$ is Euclidean self-orthogonal.
\item $C\otimes C_H$ is Hermitian self-orthogonal.
\item $C_p\otimes_p C_s$ is symplectic self-orthogonal.
\end{enumerate}
\end{theorem}
\begin{proof}
The result directly follows using Lemma~\ref{lemma:linear_tensor},
Lemma~\ref{lemma:additive_tensor}, and Theorem~\ref{theorem:product_dual}.
\end{proof}

\section{Product Codes from Cyclic Codes}
In this section we investigate the product of two cyclic codes (see
\cite[Chapter 10.4]{Bla83}, \cite[Chapter 10.2]{Bla03}).

\begin{figure}
\centerline{\footnotesize
\begin{tabular}{cc}
\begin{picture}(100,90)(-10,-10)
\multiput(0,0)(0,10){9}{\line(1,0){80}}
\multiput(0,0)(10,0){9}{\line(0,1){80}}
\multiput(15,5)(0,10){8}{\makebox(0,0){$0$}}
\multiput(25,5)(0,10){8}{\makebox(0,0){$0$}}
\multiput(35,5)(0,10){8}{\makebox(0,0){$0$}}
\multiput(75,5)(0,10){8}{\makebox(0,0){$0$}}
\multiput(5,25)(10,0){8}{\makebox(0,0){$0$}}
\multiput(5,55)(10,0){8}{\makebox(0,0){$0$}}
\put( 5,-5){\makebox(0,0){\scriptsize$0$}}
\put(15,-5){\makebox(0,0){\scriptsize$1$}}
\put(25,-5){\makebox(0,0){\scriptsize$2$}}
\put(35,-5){\makebox(0,0){\scriptsize$3$}}
\put(45,-5){\makebox(0,0){\scriptsize$4$}}
\put(55,-5){\makebox(0,0){\scriptsize$5$}}
\put(65,-5){\makebox(0,0){\scriptsize$6$}}
\put(75,-5){\makebox(0,0){\scriptsize$7$}}
\put(-10,-10){\makebox(0,0){\scriptsize$j/i$}}
\put(-5, 5){\makebox(0,0){\scriptsize$0$}}
\put(-5,15){\makebox(0,0){\scriptsize$1$}}
\put(-5,25){\makebox(0,0){\scriptsize$2$}}
\put(-5,35){\makebox(0,0){\scriptsize$3$}}
\put(-5,45){\makebox(0,0){\scriptsize$4$}}
\put(-5,55){\makebox(0,0){\scriptsize$5$}}
\put(-5,65){\makebox(0,0){\scriptsize$6$}}
\put(-5,75){\makebox(0,0){\scriptsize$7$}}
\end{picture}
&
\begin{picture}(100,90)(-10,-10)
\multiput(0,0)(0,10){9}{\line(1,0){80}}
\multiput(0,0)(10,0){9}{\line(0,1){80}}
\multiput(25,5)(10,0){3}{\makebox(0,0){$0$}}
\multiput(25,15)(10,0){3}{\makebox(0,0){$0$}}
\multiput(25,25)(10,0){3}{\makebox(0,0){$0$}}
\multiput(25,45)(10,0){3}{\makebox(0,0){$0$}}
\multiput(25,55)(10,0){3}{\makebox(0,0){$0$}}
\multiput(25,75)(10,0){3}{\makebox(0,0){$0$}}
\put( 5, 5){\makebox(0,0){$0$}}
\put(5,15){\makebox(0,0){$0$}}
\put(5,25){\makebox(0,0){$0$}}
\put(5,45){\makebox(0,0){$0$}}
\put(5,55){\makebox(0,0){$0$}}
\put(5,75){\makebox(0,0){$0$}}
\put( 5,-5){\makebox(0,0){\scriptsize$0$}}
\put(15,-5){\makebox(0,0){\scriptsize$1$}}
\put(25,-5){\makebox(0,0){\scriptsize$2$}}
\put(35,-5){\makebox(0,0){\scriptsize$3$}}
\put(45,-5){\makebox(0,0){\scriptsize$4$}}
\put(55,-5){\makebox(0,0){\scriptsize$5$}}
\put(65,-5){\makebox(0,0){\scriptsize$6$}}
\put(75,-5){\makebox(0,0){\scriptsize$7$}}
\put(-10,-10){\makebox(0,0){\scriptsize$j/i$}}
\put(-5, 5){\makebox(0,0){\scriptsize$0$}}
\put(-5,15){\makebox(0,0){\scriptsize$1$}}
\put(-5,25){\makebox(0,0){\scriptsize$2$}}
\put(-5,35){\makebox(0,0){\scriptsize$3$}}
\put(-5,45){\makebox(0,0){\scriptsize$4$}}
\put(-5,55){\makebox(0,0){\scriptsize$5$}}
\put(-5,65){\makebox(0,0){\scriptsize$6$}}
\put(-5,75){\makebox(0,0){\scriptsize$7$}}
\end{picture}
\\
a) & b)
\end{tabular}
}
\caption{Two-dimensional spectrum of a) the product of two cyclic codes
  and b) the dual code.  Blank entries may take arbitrary value.
\label{fig:spectrum}}
\end{figure}
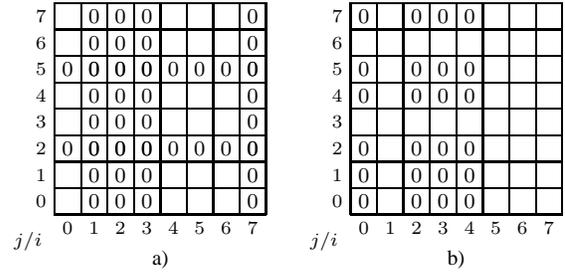

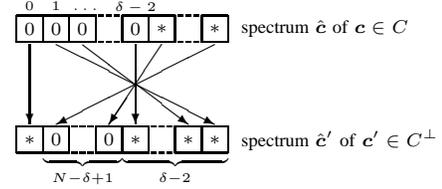
\begin{figure}
\centerline{\scriptsize
\begin{picture}(150,60)(0,-5)
\put(0,-1){\line(1,0){20}}
\multiput(0,-1)(10,0){9}{\line(0,1){10}}
\put(0,9){\line(1,0){20}}
\multiput(21,-1)(3,0){3}{\line(1,0){2}}
\multiput(21,9)(3,0){3}{\line(1,0){2}}
\put(30,-1){\line(1,0){20}}
\put(30,9){\line(1,0){20}}
\multiput(51,-1)(3,0){3}{\line(1,0){2}}
\multiput(51,9)(3,0){3}{\line(1,0){2}}
\put(60,-1){\line(1,0){20}}
\put(60,9){\line(1,0){20}}
\put(5,40){\vector(0,-1){30}}
\put(15,40){\vector(2,-1){60}}
\put(25,40){\vector(4,-3){40}}
\put(45,40){\vector(0,-1){30}}
\put(55,40){\vector(-2,-3){20}}
\put(75,40){\vector(-2,-1){60}}
\put(0,41){\line(1,0){30}}
\multiput(0,41)(10,0){9}{\line(0,1){10}}
\put(0,51){\line(1,0){30}}
\multiput(31,41)(3,0){3}{\line(1,0){2}}
\multiput(31,51)(3,0){3}{\line(1,0){2}}
\put(40,41){\line(1,0){20}}
\put(40,51){\line(1,0){20}}
\multiput(61,41)(3,0){3}{\line(1,0){2}}
\multiput(61,51)(3,0){3}{\line(1,0){2}}
\put(70,41){\line(1,0){10}}
\put(70,51){\line(1,0){10}}
\multiput(5,4)(40,0){2}{\makebox(0,0){$*$}}
\put(15,4){\makebox(0,0){$0$}}
\put(35,4){\makebox(0,0){$0$}}
\multiput(65,4)(10,0){2}{\makebox(0,0){$*$}}
\multiput(5,46)(10,0){3}{\makebox(0,0){$0$}}
\put(45,46){\makebox(0,0){$0$}}
\put(55,46){\makebox(0,0){$*$}}
\put(75,46){\makebox(0,0){$*$}}
\put(5,53){\makebox(0,0)[b]{\tiny$0$}}
\put(15,53){\makebox(0,0)[b]{\tiny$1$}}
\put(26,53){\makebox(0,0)[b]{\tiny$\ldots$}}
\put(45,52){\makebox(0,0)[b]{\tiny$\delta-2$}}
\put(85,46){\makebox(0,0)[l]{spectrum $\hat{\bm{c}}$ of $\bm{c}\in C$}}
\put(85,4){\makebox(0,0)[l]{spectrum $\hat{\bm{c}}'$ of $\bm{c}'\in C^\bot$}}
\put(25,-1){\makebox(0,0)[t]{\tiny$\underbrace{\rule{30\unitlength}{0pt}}_{N-\delta+1}$}}
\put(60,-1){\makebox(0,0)[t]{\tiny$\underbrace{\rule{40\unitlength}{0pt}}_{\delta-2}$}}
\end{picture}
}
\caption{Relation between the spectra of a Reed-Solomon code $C$ and
its dual. Positions taking arbitrary values (marked with $*$) and
positions being zero are interchanged using to the map $i\mapsto
-i\bmod (q-1)$ \cite{GrBe00}. \label{fig:RSspectrum1}}
\end{figure}

Let $C_1=[n_1,k_1]$ and $C_2=[n_2,k_2]$ be cyclic linear codes with
generator polynomials $g_1(X)$ and $g_2(Y)$.  Then $C_\pi=C_1\otimes
C_2$ is a bicyclic code generated by $g_1(X)g_2(Y)$. The codewords of
$C_\pi$ correspond to all bivariate polynomials
$c(X,Y)=i(X,Y)g_1(X)g_2(Y)$ modulo the ideal generated by $X^{n_1}-1$
and $Y^{n_2}-1$, where $i(X,Y)\in GF(q)[X,Y]$ is an arbitrary
bivariate polynomial.  The two-dimensional spectrum of $c(X,Y)$ is the
$n_1\times n_2$ matrix $(\hat{c}_{i,j})$ with entries
\begin{equation}
\hat{c}_{i,j}:=c(\alpha^i,\beta^j),
\end{equation}
where $\alpha$ and $\beta$ are primitive roots of unity of order $n_1$
and $n_2$, respectively.  The spectrum $\hat{c}$ is zero in all
vertical stripes corresponding to the roots $\alpha^i$ of $g_1(X)$ and
in all horizontal stripes corresponding to the roots $\beta^j$ of
$g_2(X)$ (see Fig.~\ref{fig:spectrum}~a)).  The generator polynomial
$h_1(X)$ of the Euclidean dual $C_1^\bot$ is the reciprocal polynomial
of $(X^{n_1}-1)/g_1(X)$.  Hence its one-dimensional spectrum is zero
at the negative of those positions where the spectrum of the code
$C_1$ takes arbitrary values ({\em cf}.~Fig.~\ref{fig:RSspectrum1}).  For
the generator polynomial $h_2(Y)$ of $C_2^\bot$ the analogous 
statement is true. Therefore the Euclidean dual code $(C_1\otimes
C_2)^\bot$ of the product code $C_1\otimes C_2$ consists of all
polynomials that are multiples of $h_1(X)$ or $h_2(Y)$.  Interchanging
the zeros and blanks in the two-dimensional spectrum of the product
code and applying the coordinate map ({\em cf}. Fig.~\ref{fig:RSspectrum1})
to both the rows and columns, we obtain the two-dimensional spectrum
of the dual code $(C_1\otimes C_2)^\bot$.

For the Hermitian dual code, we get analogous results.  As the
Hermitian inner product involves the Frobenius map $x\mapsto x^q$, the
transformation on the coordinates now reads $i\mapsto -qi \bmod n_j$.

\begin{figure}
\centerline{\footnotesize\unitlength0.9\unitlength
\begin{tabular}{cc}
\begin{picture}(130,110)(-20,-10)
\multiput( 0, 0)(0,10){4}{\line(1,0){30}}
\multiput(30, 0)(0,10){4}{\multiput(2.5,0)(6,0){2}{\line(1,0){3}}}
\multiput(45, 0)(0,10){4}{\line(1,0){20}}
\multiput(65, 0)(0,10){4}{\multiput(2.5,0)(6,0){4}{\line(1,0){3}}}
\multiput(90, 0)(0,10){4}{\line(1,0){10}}
\multiput( 0,55)(0,10){3}{\line(1,0){30}}
\multiput(30,55)(0,10){3}{\multiput(2.5,0)(6,0){2}{\line(1,0){3}}}
\multiput(45,55)(0,10){3}{\line(1,0){20}}
\multiput(65,55)(0,10){3}{\multiput(2.5,0)(6,0){4}{\line(1,0){3}}}
\multiput(90,55)(0,10){3}{\line(1,0){10}}
\multiput( 0,90)(0,10){2}{\line(1,0){30}}
\multiput(30,90)(0,10){2}{\multiput(2.5,0)(6,0){2}{\line(1,0){3}}}
\multiput(45,90)(0,10){2}{\line(1,0){20}}
\multiput(65,90)(0,10){2}{\multiput(2.5,0)(6,0){4}{\line(1,0){3}}}
\multiput(90,90)(0,10){2}{\line(1,0){10}}
\multiput( 0, 0)(10,0){4}{\line(0,1){30}}
\multiput( 0,30)(10,0){4}{\multiput(0,2.5)(0,6){4}{\line(0,1){3}}}
\multiput( 0,55)(10,0){4}{\line(0,1){20}}
\multiput( 0,75)(10,0){4}{\multiput(0,2.5)(0,6){2}{\line(0,1){3}}}
\multiput( 0,90)(10,0){4}{\line(0,1){10}}
\multiput(45, 0)(10,0){3}{\line(0,1){30}}
\multiput(45,30)(10,0){3}{\multiput(0,2.5)(0,6){4}{\line(0,1){3}}}
\multiput(45,55)(10,0){3}{\line(0,1){20}}
\multiput(45,75)(10,0){3}{\multiput(0,2.5)(0,6){2}{\line(0,1){3}}}
\multiput(45,90)(10,0){3}{\line(0,1){10}}
\multiput(90, 0)(10,0){2}{\line(0,1){30}}
\multiput(90,30)(10,0){2}{\multiput(0,2.5)(0,6){4}{\line(0,1){3}}}
\multiput(90,55)(10,0){2}{\line(0,1){20}}
\multiput(90,75)(10,0){2}{\multiput(0,2.5)(0,6){2}{\line(0,1){3}}}
\multiput(90,90)(10,0){2}{\line(0,1){10}}
\multiput( 5, 5)(0,10){3}{\multiput(0,0)(10,0){3}{\makebox(0,0){$0$}}}
\multiput(50, 5)(0,10){3}{\multiput(0,0)(10,0){2}{\makebox(0,0){$0$}}}
\multiput(95, 5)(0,10){3}{\makebox(0,0){$0$}}
\multiput( 5,60)(0,10){2}{\multiput(0,0)(10,0){3}{\makebox(0,0){$0$}}}
\multiput( 5,95)(10,0){3}{\makebox(0,0){$0$}}
\multiput( 5,60)(10,0){3}{\makebox(0,0){$0$}}
\multiput(50,60)(0,10){2}{\makebox(0,0){$0$}}
\put(50,95){\makebox(0,0){$0$}}
\put(60,60){\makebox(0,0){$0$}}
\put(95,60){\makebox(0,0){$0$}}
\put( 5,-7){\makebox(0,0){\scriptsize$0$}}
\put(15,-7){\makebox(0,0){\scriptsize$1$}}
\put(25,-7){\makebox(0,0){\scriptsize$2$}}
\put(50,-7){\makebox(0,0){\scriptsize$\delta_1{-}2$}}
\put(95,-7){\makebox(0,0){\scriptsize$q{-}2$}}
\put(-10,-10){\makebox(0,0){\scriptsize$j/i$}}
\put(-3, 5){\makebox(0,0)[r]{\scriptsize$0$}}
\put(-3,15){\makebox(0,0)[r]{\scriptsize$1$}}
\put(-3,25){\makebox(0,0)[r]{\scriptsize$2$}}
\put(-3,60){\makebox(0,0)[r]{\scriptsize$\delta_2{-}2$}}
\put(-3,95){\makebox(0,0)[r]{\scriptsize$q{-}2$}}
\end{picture}
&
\begin{picture}(130,110)(-20,-10)
\multiput( 0, 0)(0,10){4}{\line(1,0){30}}
\multiput(30, 0)(0,10){4}{\multiput(2.5,0)(6,0){4}{\line(1,0){3}}}
\multiput(55, 0)(0,10){4}{\line(1,0){20}}
\multiput(75, 0)(0,10){4}{\multiput(2.5,0)(6,0){2}{\line(1,0){3}}}
\multiput(90, 0)(0,10){4}{\line(1,0){10}}
\multiput( 0,45)(0,10){3}{\line(1,0){30}}
\multiput(30,45)(0,10){3}{\multiput(2.5,0)(6,0){4}{\line(1,0){3}}}
\multiput(55,45)(0,10){3}{\line(1,0){20}}
\multiput(75,45)(0,10){3}{\multiput(2.5,0)(6,0){2}{\line(1,0){3}}}
\multiput(90,45)(0,10){3}{\line(1,0){10}}
\multiput( 0,90)(0,10){2}{\line(1,0){30}}
\multiput(30,90)(0,10){2}{\multiput(2.5,0)(6,0){4}{\line(1,0){3}}}
\multiput(55,90)(0,10){2}{\line(1,0){20}}
\multiput(75,90)(0,10){2}{\multiput(2.5,0)(6,0){2}{\line(1,0){3}}}
\multiput(90,90)(0,10){2}{\line(1,0){10}}
\multiput( 0, 0)(10,0){4}{\line(0,1){30}}
\multiput( 0,30)(10,0){4}{\multiput(0,2.5)(0,6){2}{\line(0,1){3}}}
\multiput( 0,45)(10,0){4}{\line(0,1){20}}
\multiput( 0,65)(10,0){4}{\multiput(0,2.5)(0,6){4}{\line(0,1){3}}}
\multiput( 0,90)(10,0){4}{\line(0,1){10}}
\multiput(55, 0)(10,0){3}{\line(0,1){30}}
\multiput(55,30)(10,0){3}{\multiput(0,2.5)(0,6){2}{\line(0,1){3}}}
\multiput(55,45)(10,0){3}{\line(0,1){20}}
\multiput(55,65)(10,0){3}{\multiput(0,2.5)(0,6){4}{\line(0,1){3}}}
\multiput(55,90)(10,0){3}{\line(0,1){10}}
\multiput(90, 0)(10,0){2}{\line(0,1){30}}
\multiput(90,30)(10,0){2}{\multiput(0,2.5)(0,6){2}{\line(0,1){3}}}
\multiput(90,45)(10,0){2}{\line(0,1){20}}
\multiput(90,65)(10,0){2}{\multiput(0,2.5)(0,6){4}{\line(0,1){3}}}
\multiput(90,90)(10,0){2}{\line(0,1){10}}
\multiput(15,15)(0,10){2}{\multiput(0,0)(10,0){2}{\makebox(0,0){$0$}}}
\multiput(60,15)(0,10){2}{\makebox(0,0){$0$}}
\multiput(15,50)(10,0){2}{\makebox(0,0){$0$}}
\put(60,50){\makebox(0,0){$0$}}
\put( 5,-7){\makebox(0,0){\scriptsize$0$}}
\put(15,-7){\makebox(0,0){\scriptsize$1$}}
\put(25,-7){\makebox(0,0){\scriptsize$2$}}
\put(60,-7){\makebox(0,0){\scriptsize$\delta'_1{-}1$}}
\put(95,-7){\makebox(0,0){\scriptsize$q{-}2$}}
\put(-10,-10){\makebox(0,0){\scriptsize$j/i$}}
\put(-3, 5){\makebox(0,0)[r]{\scriptsize$0$}}
\put(-3,15){\makebox(0,0)[r]{\scriptsize$1$}}
\put(-3,25){\makebox(0,0)[r]{\scriptsize$2$}}
\put(-3,50){\makebox(0,0)[r]{\scriptsize$\delta'_2{-}1$}}
\put(-5,95){\makebox(0,0)[r]{\scriptsize$q{-}2$}}
\end{picture}
\\
a) & b)
\end{tabular}
}
\caption{Two-dimensional spectrum of a) the product of two Reed-Solomon
  codes $C_1$ and $C_2$ with minimum distance $\delta_1$ and $\delta_2$, and b) the
  dual code,  where $\delta'_1$ and $\delta'_2$ denote minimum
  distance of the dual codes $C_1^\bot$ and $C_2^\bot$.
\label{fig:RSspectrum2}}
\end{figure}
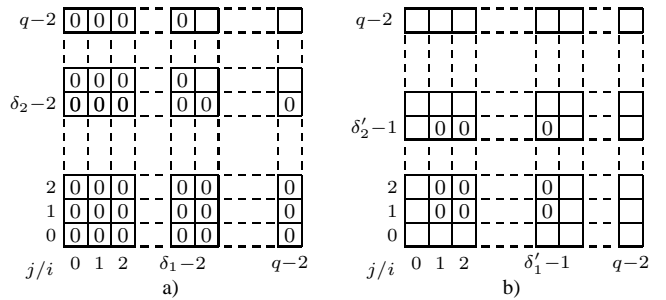

For Reed-Solomon codes, the picture simplifies.  The two-dimensional
spectrum of the product of two Reed-Solomon codes with minimum
distance $\delta_1$ and $\delta_2$ corresponds to a vertical stripe of
zeros of width $\delta_1-1$ and a horizontal stripe of height
$\delta_2-1$.  Without loss of generality, the stripes can be shifted
such that the rectangle of arbitrary values is in the upper right
corner (see Fig.~\ref{fig:RSspectrum2} a).  Then for the dual code,
the spectrum is zero in a rectangle (see Fig.~\ref{fig:RSspectrum2} b)
whose width and height is determined by the dual distances
$(q-\delta_1)$ and $(q-\delta_1)$.  Using the BCH-like lower bound for
bicyclic codes (see \cite[p. 320]{Bla03}), we conclude that the
minimum distance of the dual of the product code is
$\min(q-\delta_1,q-\delta_2)$.  In summary, we get the following
theorem:
\begin{theorem}\label{theorem:RS_prod}
The product code of two Reed-Solomon codes
$C_1=[q-1,q-\delta_1,\delta_1]_q$ and
$C_2=[q-1,q-\delta_2,\delta_2]_q$ over $GF(q)$ is
\begin{equation}
C_1\otimes C_2=[(q-1)^2,(q-\delta_1)(q-\delta_2),\delta_1\delta_2]_q.
\end{equation}
The Euclidean dual code $(C_1\otimes C_2)^\bot=[(q-1)^2,K^\bot,d^\bot]_q$
has parameters
\begin{alignat*}{2}
K^\bot&=q(d_1+d_2-2)-d_1 d_2 +1\\
d^\bot&=\min(q-\delta_1,q-\delta_2).
\end{alignat*}
Moreover, the product code is self-orthogonal if $C_1$ or $C_2$ is self-orthogonal.
\end{theorem}
Note that the result is still true when replacing the Reed-Solomon
code over $GF(q)$ of length $(q-1)$ by a cyclic code $C=[n,k,d]_q$ with
generator polynomial $g(X)=\prod_{i=0}^{d-2}(X-\alpha^i)$ where $n$ is
a divisor of $q-1$ and $\alpha$ is a primitive $n$-th root of unity.

\section{Quantum Codes from Product Codes}
\subsection{Quantum Block Codes}
In the previous section we have seen that the product of a
self-orthogonal Reed-Solomon code with an arbitrary Reed-Solomon codes
yields a self-orthogonal product code.  Using Lemma~\ref{lemma:CSS},
we can construct quantum error-correcting codes.
\begin{theorem}\label{theorem:RS_prod_QECC}
Let $C_1=[q-1,\mu_1,q-\mu_1]_q$ and $C_2=[q-1,\mu_2,q-\mu_2]_q$ be
Reed-Solomon codes where $\mu_1<(q-1)/2$.  Then a quantum
error-correcting code
\begin{equation}\label{eq:RS_prod_QECC}
{\cal C}=\QECC((q-1)^2,(q-1)^2-2\mu_1\mu_2,{1+\min(\mu_1,\mu_2)},q)
\end{equation}
exists.
\end{theorem}
\begin{proof}
For $\mu_1<(q-1)/2$, the code $C_1$ is Euclidean self-orthogonal
\cite{GBR04}. The dual distance of $C_1$ and $C_2$ is $\mu_1+1$ and
$\mu_2+1$, respectively.  By Theorem~\ref{theorem:RS_prod}, the
product code $C_\pi=C_1\otimes
C_2=[(q-1)^2,\mu_1\mu_2,(q-\mu_1)(q-\mu_2)]_q$ is self-orthogonal.
Its Euclidean dual has parameters
$C_\pi^\bot=[(q-1)^2,(q-1)^2-\mu_1\mu_2,1+\min(\mu_1,\mu_2)]_q$.
Hence by Lemma~\ref{lemma:CSS} a QECC with the parameters given in
eq.~(\ref{eq:RS_prod_QECC}) exists.
\end{proof}
Note that from $C_1$ and $C_2$ (provided $\mu_2<(q-1)/2$), one can
construct optimal QECCs with parameters $\QECC(q-1,q-2\mu-1,\mu+1,q)$ (see
\cite{GBR04}).  The product of the rates of these codes is
$$
\left(1-\frac{2\mu_1}{q-1}\right)\left(1-\frac{2\mu_2}{q-1}\right)
=1-\frac{2(\mu_1+\mu_2)}{q-1}+\frac{4\mu_1\mu_2}{(q-1)^2}
$$
The rate of the code of Theorem~\ref{theorem:RS_prod_QECC} is
$$
1-\frac{2\mu_1\mu_2}{(q-1)^2}.
$$
If we choose $\mu_1=\mu_2$, we will obtain a QECC of squared length
and the same minimum distance, but higher rate provided
$\mu_1=\mu_2<2(q-1)/3$.

Note that we can obtain good QECCs by this construction using other
codes than Reed-Solomon codes.  Let $C=[5,2,4]_4$ be the Hermitian
dual of the quaternary Hamming code.  Using $C\subseteq
C^*=[5,3,3]_4$, an optimal QECC ${\cal C}=\QECC(5,1,3,2)$ can be
constructed.  The code $C$ is not a Reed-Solomon code, but its
spectrum fulfills the conditions for Theorem~\ref{theorem:RS_prod}.
Hence the product of $C$ with itself is a Hermitian self-orthogonal
code $C\otimes C=[25,4,16]_4\subseteq (C\otimes C)^*=[25,21,3]_4$.
This yields a QECC ${\cal C}^{(2)}=\QECC(25,17,3,2)$, whose rate is
more than three times higher than that of ${\cal C}$.

The product code of $C$, considered as additive code, with
the binary simplex code $C_1=[3,2,2]_2$ is an additive code
$C_2:=C_1\otimes_p C=(15,2^8,8)_2$ which is contained in its symplectic
dual $C^\star=(15,2^{22},3)_2$.  Hence we obtain a QECC ${\cal C}_\pi=\QECC(15,7,3,2)$.

\section{Quantum Convolutional Codes}
Following \cite{OlTi04}, an $(n,k,m)$ quantum convolutional code can
be described in terms of a semi-infinite stabilizer matrix $S$.  The
matrix $S$ has a block band
structure where each block $M$ has size $(n-k)\times (n+m)$.  All
blocks are equal.  In the second block, the matrix $M$ is shifted by
$n$ columns, hence any two consecutive blocks overlap in $m$
positions.  The general structure of the matrix is as follows:
$$
S:=\left(\rule[-10mm]{0pt}{17ex}\right.
\begin{array}{l}
\overbrace{\rule{17mm}{0pt}}^{n}%
\overbrace{\rule{10mm}{0pt}}^{m}\\
\framebox[28mm]{\rule[-1ex]{0pt}{3ex}{$M$}}\left.\rule{0pt}{2.7ex}\right\}n-k\\
\rule{17mm}{0pt}\framebox[28mm]{\rule[-1ex]{0pt}{3ex}{$M$}}\\
\rule{34mm}{0pt}\framebox[28mm]{\rule[-1ex]{0pt}{3ex}{$M$}}\\
\rule{51mm}{0pt}\ddots\\
\rule{0pt}{1ex}
\end{array}
\left.\rule[-10mm]{0pt}{17ex}\right)
$$

The classical convolutional code generated by $S$ must be
self-orthogonal with respect to some of the inner products of
Section~\ref{sec:inner_products}.
The quantum product codes constructed in the previous section
naturally lend themselves to convolutional codes because of the
following observation. Let $M=G^{(1)} \otimes G^{(2)}$ be the
generator matrix of $C_1 \otimes C_2$ as in
eq.~(\ref{eq:block_structure}). Assume that $m=t n_2$ is a multiple of
$n_2$, the length of $C_2$. Since $C_2$ is self-orthogonal, we have
that the submatrix of $M$ which consists of the last $m$ columns of
$M$ is orthogonal to the submatrix which consists of the first $m$
columns of $M$. Hence, we obtain a semi-infinite stabilizer matrix $S$
by iterative shifting of the block $M$ by $n_1 n_2-m=(n_1-t)n_2$
positions. 

To give an example, we let $C=[7,3,4]_2$ be the Euclidean dual of the
binary Hamming code. Using $C\subseteq C^\bot=[7,4,3]_2$, a QECC
${\cal C}=\QECC(7,1,3,2)$ can be constructed.  The product code of $C$
with itself is a code $C_\pi=C\otimes C=[49,9,16]_2$ which is
contained in its dual $C_\pi^\bot=[49,40,3]_2$.  Hence we obtain a
QECC ${\cal C}_\pi=\QECC(49,31,3,2)$. The possible parameters for
quantum convolutional codes obtained from the product code $C_\pi$ by
the CSS construction ({\em i.\,e.}, by considering the generator matrix
$C_\pi\otimes GF(4)$) are $(49-m,31,m)$, $m=7,14$. The free distance
of these codes is $3$.  Using tail-biting with $N\ge 2$ blocks and
$m=7$ (see \cite{FoGu05}) we obtain QECCs $\QECC(42N,24N,d,2)$.  Using
\Magma{} \cite{Magma} we compute $d=3$.

From the product code
${\cal C}_\pi=\QECC(15,7,3,2)$ described above we can obtain a quantum
convolutional code with parameters $(10,7,5)$, {\em i.\,e.}, we choose
$m=5$.

\begin{figure}
{\footnotesize
$$\arraycolsep0.2\arraycolsep
\left(
\begin{array}{ccccccccc}
g_{11}^{(1)}G^{(2)} & g_{12}^{(1)}G^{(2)} & \ldots &g_{1,n_1}^{(1)} G^{(2)}\\
g_{21}^{(1)}G^{(2)} & g_{22}^{(1)}G^{(2)} & \ldots &g_{2,n_1}^{(1)} G^{(2)}\\
\vdots &\vdots&\ddots&\vdots\\
g_{k_1,1}^{(1)}G^{(2)} & g_{k_1,2}^{(1)}G^{(2)} & \ldots & g_{k_1,n_1}^{(1)} G^{(2)}\\[2ex]
&&&g_{11}^{(1)}G^{(2)} & g_{12}^{(1)}G^{(2)} & \ldots &g_{1,n_1}^{(1)} G^{(2)}\\
&&&g_{21}^{(1)}G^{(2)} & g_{22}^{(1)}G^{(2)} & \ldots &g_{2,n_1}^{(1)} G^{(2)}\\
&&&\vdots &\vdots&\ddots&\vdots\\
&&&g_{k_1,1}^{(1)}G^{(2)} & g_{k_1,2}^{(1)}G^{(2)} & \ldots & g_{k_1,n_1}^{(1)} G^{(2)}\\
&&&&&&\vdots&\ddots
\end{array}
\right)
$$
}
{\footnotesize
$$\arraycolsep0.2\arraycolsep
{}=\left(
\begin{array}{cccccccccc}
g_{11}^{(1)} & g_{12}^{(1)} & \ldots &g_{1,n_1}^{(1)} \\
g_{21}^{(1)} & g_{22}^{(1)} & \ldots &g_{2,n_1}^{(1)} \\
\vdots &\vdots&\ddots&\vdots\\
g_{k_1,1}^{(1)} & g_{k_1,2}^{(1)} & \ldots & g_{k_1,n_1}^{(1)} \\[2ex]
&&&g_{11}^{(1)} & g_{12}^{(1)} & \ldots &g_{1,n_1}^{(1)} \\
&&&g_{21}^{(1)} & g_{22}^{(1)} & \ldots &g_{2,n_1}^{(1)} \\
&&&\vdots &\vdots&\ddots&\vdots\\
&&&g_{k_1,1}^{(1)} & g_{k_1,2}^{(1)} & \ldots & g_{k_1,n_1}^{(1)} \\
&&&&&&\vdots&\ddots
\end{array}
\right)\otimes G^{(2)}
$$
}
\caption{Tensor product decomposition of the semi-infinite band matrix
  derived from the generator matrix of a product code (here shown for
  $t=1$).\label{fig:tensor_band}}
\end{figure}

If the matrix $M$ defining the semi-infinite band matrix $S$ is the
generator matrix $G^{(1)}\otimes G^{(2)}$ of a product code, the
matrix $S$ itself can be decomposed as a tensor product
$S=S^{(1)}\otimes G^{(2)}$, provided the overlap $m$ is a multiple of
the length $n_2$ of the second code, {\em i.\,e.}, $m=t n_2$ (see
Fig.~\ref{fig:tensor_band}).  The matrix $S^{(1)}$ is a semi-infinite
band matrix with $M^{(1)}=G^{(1)}$ and overlap $t$.  From
Theorem~\ref{theorem:tensor_selfdual} it follows that the product code
is self-orthogonal if $C_2$ is self-orthogonal.  Hence we get the
following construction:
\begin{theorem}
Let $C_1$ be a classical convolutional code.  Furthermore, let $C_2$
be a self-orthogonal code.  Then the product code $C_1\otimes C_2$
defines a quantum convolutional code, provided at least one of the
following holds:
\def\theenumi{(\roman{enumi}}
\begin{enumerate}
\item Both $C_1$ and $C_2$ are linear over $GF(q)$ and $C_2$ is Euclidean
  self-orthogonal.
\item Both $C_1$ and $C_2$ are linear over $GF(q^2)$ and $C_2$ is Hermitian
  self-orthogonal.
\item $C_1$ is linear of $GF(p)$ and $C_2$ is a symplectic
  self-orthogonal code over $GF(p^\ell)$.
\end{enumerate}
\end{theorem}
\enlargethispage{-0.86in}

\section{Conclusion}
The construction of new examples of quantum convolutional codes is a
challenging task and rises several questions: what is a general
framework to describe such codes, how can they be constructed, and
what are the figures of merit to compare the performance of such
codes? While the first of these questions has been answered in a
satisfying way at least for convolutional stabilizer codes in
\cite{OlTi04}, the other two questions are open (but see {\em e.\,g.}
\cite{AlPa04,FoGu05,OlTi03}). In this paper we have contributed to the
second question by establishing a connection between product codes and
convolutional codes.  We have shown that the dual distance of product
codes can be bounded from below which allows to obtain quantum codes
for which the minimum distance is at least as large as the smaller of
the minimum distances of the factors.

Concerning the third question currently not much is known, {\em e.\,g.}, the
significance of notions such as {\em free distance} which are useful
for classical convolutional codes to the quantum case has yet to be
investigated. 


\section*{Acknowledgment}
% optional entry into table of contents (if used)
%\addcontentsline{toc}{section}{Acknowledgment}
This work was carried out while the second author was visiting IAKS.
M.R. also acknowledges support by the Institute of Quantum Computing,
University of Waterloo.  Funding by {\em Deutsche
Forschungsgemeinschaft (DFG), Schwerpunktprogramm
Quanten-Informationsverarbeitung (SPP 1078), Projekt AQUA (Be~887/13)}
is acknowledged as well.


\begin{thebibliography}{10}
\providecommand{\url}[1]{#1}
\csname url@rmstyle\endcsname
\providecommand{\newblock}{\relax}
\providecommand{\bibinfo}[2]{#2}
\providecommand\BIBentrySTDinterwordspacing{\spaceskip=0pt\relax}
\providecommand\BIBentryALTinterwordstretchfactor{4}
\providecommand\BIBentryALTinterwordspacing{\spaceskip=\fontdimen2\font plus
\BIBentryALTinterwordstretchfactor\fontdimen3\font minus
  \fontdimen4\font\relax}
\providecommand\BIBforeignlanguage[2]{{%
\expandafter\ifx\csname l@#1\endcsname\relax
\typeout{** WARNING: IEEEtran.bst: No hyphenation pattern has been}%
\typeout{** loaded for the language `#1'. Using the pattern for}%
\typeout{** the default language instead.}%
\else
\language=\csname l@#1\endcsname
\fi
#2}}

\bibitem{AsKn01}
A.~Ashikhmin and E.~Knill, ``{Nonbinary quantum stabilizer codes},'' \emph{IEEE
  Transactions on Information Theory}, vol.~47, no.~7, pp. 3065--3072, Nov.
  2001, {Preprint quant-ph/0005008}.

\bibitem{Bla83}
R.~E. Blahut, \emph{Theory and Practice of Error Control Codes}.\hskip 1em plus
  0.5em minus 0.4em\relax Reading: Addison-Wesley, 1983.

\bibitem{Bla03}
------, \emph{Algebraic Codes for Data Transmission}.\hskip 1em plus 0.5em
  minus 0.4em\relax Cambridge: Cambridge University Press, 2003.

\bibitem{Magma}
W.~Bosma, J.~J. Cannon, and C.~Playoust, ``{The Magma Algebra System I: The
  User Language},'' \emph{Journal of Symbolic Computation}, vol.~24, no. 3--4,
  pp. 235--266, 1997.

\bibitem{CaSh96}
A.~R. Calderbank and P.~W. Shor, ``{Good quantum error-correcting codes
  exist},'' \emph{Physical Review~A}, vol.~54, no.~2, pp. 1098--1105, Aug.
  1996, preprint quant-ph/9512032.

\bibitem{ChNg73}
R.~T. Chien and S.~W. Ng, ``{Dual Product Codes for Correction of Multiple
  Low-Density Burst Errors},'' \emph{IEEE Transactions on Information Theory},
  vol.~19, no.~5, pp. 672--677, Sept. 1973.

\bibitem{AlPa04}
A.~C.~A. de~Almeida and R.~Palazzo, Jr., ``{A Concatenated $[(4,1,3)]$ Quantum
  Convolutional Code},'' in \emph{2004 IEEE Information Theory Workshop}, San
  Antonio, TX, Oct. 2004.

\bibitem{FoGu05} G.~D. Forney, Jr. and S.~Guha, ``{Simple rate-$1/3$
    convolutional and tail-biting quantum error-correcting codes},''
  in \emph{{Proc. ISIT'05, Adelaide, Australia}}, 2005, pp.  1028--1032.


\bibitem{GrBe00}
M.~Grassl and Th.~Beth, ``{Cyclic quantum error-correcting codes and quantum
  shift registers},'' \emph{Proceedings of the Royal Society London A}, vol.
  456, no. 2003, pp. 2689--2706, Nov. 2000, preprint quant-ph/9910061.

\bibitem{GBR04}
M.~Grassl, Th.~Beth, and M.~R{\"o}tteler, ``{On Optimal Quantum Codes},''
  \emph{International Journal of Quantum Information}, vol.~2, no.~1, pp.
  55--64, 2004, preprint quant-ph/0312164.

\bibitem{MS77}
F.~J. MacWilliams and N.~J.~A. Sloane, \emph{The Theory of Error--Correcting
  Codes}.\hskip 1em plus 0.5em minus 0.4em\relax Amsterdam: North--Holland,
  1977.

\bibitem{OlTi03}
H.~Ollivier and J.-P. Tillich, ``Description of a quantum convolutional code,''
  \emph{Physical Review Letters}, vol.~91, 177902, Oct. 24 2003.

\bibitem{OlTi04}
--------, ``Quantum convolutional codes: fundamentals,'' Nov. 2004, preprint
  quant-ph/0401134.

\bibitem{Ste96:error}
A.~M. Steane, ``{Error Correcting Codes in Quantum Theory},'' \emph{Physical
  Review Letters}, vol.~77, no.~5, pp. 793--797, 29.~July 1996.

\bibitem{Wol65}
J.~K. Wolf, ``{On Codes Derivable from the Tensor Product of Check Matrices},''
  \emph{IEEE Transactions on Information Theory}, vol.~11, no.~2, pp. 281--284,
  Apr. 1965.

\end{thebibliography}
\end{document}